\documentclass[twocol]{ametsocV5}

\usepackage{amsmath,amsfonts,amssymb,bm}

\renewcommand{\vec}[1]{\mathbf{#1}}

\title{On the connection between intermittency and dissipation in ocean turbulence}

\authors{Jordi Isern-Fontanet \correspondingauthor{Jordi Isern-Fontanet, jisern@icm.csic.es}, Antonio Turiel}

\affiliation{Institut de Ci\`encies del Mar (CISC), Barcelona, EU}

\abstract{The multifractal theory of turbulence is used to investigate the energy cascade in the Northwestern Atlantic ocean. The statistics of singularity exponents of velocity gradients computed from in situ measurements are used to show that the anomalous scaling of the velocity structure functions at depths between 50 ad 500 m has a linear dependence on the exponent characterizing the strongest velocity gradient, with a slope that decreases with depth. Since the distribution of exponents is asymmetric about the mode at all depths, we use an infinitely divisible asymmetric model of the energy cascade, the log-Poisson model, to derive the functional dependence of the anomalous scaling with dissipation. Using this model we can interpret the vertical change of the linear slope as a change in the energy cascade.}

\begin{document}

\maketitle

%
%
%


\section{Introduction}

Predicting the evolution of a turbulent flow such as the ocean motion remains a long-standing problem in physics. Beyond the difficulties of developing a closed theory from fundamental laws, the range of interacting scales involved in oceanic motions, i.e. between $10^{-3}$m and  $10^{7}$m, is too large to be resolved by ocean and climate numerical models \citep{Ferrari.2008, Poje.2014, Su.2018}. As a consequence, some kind of approximation or parametrization of the unresolved scales has to be implemented to correctly reproduce the flux of energy between scales \citep{Pearson.2017, Dubrulle.2019}.  But, although  in three-dimensional turbulence energy is known to cascade from scales at which it is injected towards smaller scales where it is utterly converted to heat, the details of such cascade in the ocean are rather complex and still poorly understood \citep{Muller.2005, McWilliams.2016, Renault.2019}. This hinders its correct parametrization and becomes a source of errors in ocean and climate models \citep{Flato.2013}. 

In the statistical equilibrium, the global characteristics of the three-dimensional energy cascade are controlled by energy dissipation at small scales and its statistical properties completely characterize the whole statistics of the velocity field. As a consequence, the spotty nature of dissipation imprints an intermittent behavior to velocities. 
The statistical impact of intermittence is the heavy tails of velocity Probability Density Functions (PDF) and the so-called anomalous scaling of the velocity structure functions, i.e. the moments of velocity differences. These effects could not be explained by Kolmogorov 1941 theory and required some refinements  \citep{Kolmogorov.1962, Oboukhov.1962}. The multifractal theory of fully developed turbulence \citep{Benzi.1984,Parisi.1985} allowed to fill this theoretical gap. Multifractal theory assumes that dissipation and velocity are defined on multiple sets of points with a fractal dimension, each one associated to a singularity exponent in a continuum spectrum. The cornerstone of this theory is the singularity spectrum, which links the geometry of the flow (the fractal dimensions of the fractal components) with the scaling properties of the variable and flow statistics \citep{Boffetta.2008}. Although the validity of the multifractal framework has been assessed for the oceans \citep{Lovejoy.2001, Turiel.2005, Isern.2007, Seuront.2014}, the choice of the relevant singularity spectrum model \citep{Kolmogorov.1962, Schertzer.1987, She.1995} and its parameters depends on unknown underlying physical processes and remains an open question in turbulence \citep{Dubrulle.2019}. 

Here, we investigate the connection between intermittency and singularity spectra in the upper layers of the ocean, and their dependence on dissipation. Our approach exploits recent advances in computational mathematics, which have led to a fast and robust algorithm for computing singularity exponents \citep{Pont.2013}. Singularity spectra are thus obtained from the singularity exponents derived from measurements by a ship-of-opportunity in the North-Western Atlantic ocean \citep{Rossby.2010, Callies.2015}. This allows us to avoid many numerical problems (as for instance data scarcity) \citep{Turiel.2006} and to identify the location of the most dissipative structures. Theoretical predictions from  energy cascade models can be tested and used to provide a physical interpretation of the velocity intermittence. 

The paper is organized as follows. Section \ref{sec:theory} provides a coherent view of the multifractal theory of turbulence, section \ref{sec:data} and \ref{sec:methods} are about data and procedures, section \ref{sec:obs} describes the approach used to analyze the data and outlines the key findings and section \ref{sec:model} presents a theoretical justification of our findings. The general discussion of our study can be found in section \ref{sec:discussion} and conclusions in the final section (section \ref{sec:conclusion}). 

\section{The multifractal theory of turbulence}
\label{sec:theory}

\subsection{Energy budget of coarse-grained velocities}

The momentum equation for the Bousinesq approximation is given by 
\begin{equation}
\begin{split}
\frac{\partial \vec{u}}{\partial t} + \vec{u}\cdot\nabla \vec{u}  &=-2\vec{\Omega}\times\vec{u} - \frac{1}{\rho_0}\nabla p  + b\vec{e}_z \\
                                                                                             &+\nu\nabla^2\vec{u} + \vec{F},
\end{split}
\label{eq:Bousinesq}
\end{equation}
where $\vec{u}(\vec{x})$ is the velocity field; $p(\vec{x})$ the pressure; $\vec{\Omega}$ the Earth rotation; $\nu$ the viscosity; $b(\vec{x})$ the buoyancy; $\vec{F}(\vec{x})$ an external forcing; $\rho_0$ the reference density and $\vec{e}_z$ is the vertical unit vector \citep{Vallis.2006}\footnote{The time dependence of all variables has been dropped to alleviate the notation.}.  Following \citet{Eyink.2005}, the coarse-grained momentum equation at scale $r$ is derived by applying a low-pass filter to velocities,
\begin{equation}
\bar{\vec{u}}_r (\vec{x})\equiv\int_{\mathbb{R}^d} d\vec{x}'\frac{1}{r^d}G\left(\frac{\vec{x}}{r}\right)\vec{u}(\vec{x}+\vec{x}'),
\label{eq:coarse-grained}
\end{equation}
and the other variables, where 
$d$ is the geometric dimension of the space. The filtering function $G(\vec{x})$ is taken to be positive, normalized and $G(\vec{x}) \rightarrow 0$ rapidly for $|\vec{x}|\rightarrow\infty$. Moreover, the above filtering is  such that it commutes with spatial derivatives, e.g.
\begin{equation}
\nabla\cdot\bar{\vec{u}}_r = \overline{(\nabla\cdot\vec{u})}_r.
\label{eq:commute}
\end{equation}
Then, the resulting coarse-grained momentum equation is
\begin{equation}
\begin{split}
\frac{\partial \bar{\vec{u}}_r }{\partial t} + \bar{\vec{u}}_r \cdot\nabla \bar{\vec{u}}_r    &=-2\vec{\Omega}\times\bar{\vec{u}}_r - \frac{1}{\rho_0}\nabla \bar{p}_r   + \bar{b}_r \vec{e}_z \\
                                                                                             &+\nu\nabla^2\bar{\vec{u}}_r  + \bar{\vec{F}}_r  - \nabla \cdot \mathcal{T}_r.
\end{split}
\label{eq:Bousinesq-coarse-grained}
\end{equation}
The new term  $\mathcal{T}_r(\vec{x})$ in the filtered equations is the turbulent stress tensor defined by
\begin{equation}
\mathcal{T}_r(\vec{x})\equiv\overline{\vec{u}\vec{u}}_r-\bar{\vec{u}}_r\bar{\vec{u}}_r,
\label{eq:T-def}
\end{equation}
which is responsible of coupling large scales ($>r$) with small scales ($<r$). The difficulties to write  $\mathcal{T}_r(\vec{x})$  as a function of $ \bar{\vec{u}}_r$ is what is called the closure problem in turbulence.

The equation for the conservation of energy of the large scale flow  $|\bar{\vec{u}}_r|^2/ 2$ is derived  multiplying  equation \ref{eq:Bousinesq-coarse-grained} by $\bar{\vec{u}}_r(\vec{x})$, which gives
\begin{equation}
\frac{1}{2}\frac{\partial |\bar{\vec{u}}_r|^2}{\partial t}+\nabla\cdot \vec{J}_r = - \Pi_r-\varepsilon_r+ \bar{b}_r \vec{e}_z\cdot\bar{\vec{u}}_r+\bar{\vec{F}}_r\cdot\bar{\vec{u}}_r.
\label{eq:ke-coarse-grained}
\end{equation}
The second term  in the left-hand side of equation \ref{eq:ke-coarse-grained} is the spatial transport of energy, with
\begin{equation}
\vec{J}_r (\vec{x})\equiv \frac{1}{2}|\bar{\vec{u}}_r|^2\bar{\vec{u}}_r+\frac{1}{\rho_0}\bar{p}_r\bar{\vec{u}}_r-\frac{\nu}{2}\nabla|\bar{\vec{u}}_r|^2+  \bar{\vec{u}}_r\cdot \mathcal{T}_r.
\end{equation}
In the right hand side of equation \ref{eq:ke-coarse-grained}, the first term is the local flux of energy between scales
\begin{equation}
\Pi_r(\vec{x})\equiv-\bar{\mathcal{S}}_r:\mathcal{T}_r,
\label{eq:Pi-def}
\end{equation}
where $\bar{\mathcal{S}}_r$ is the large scale strain tensor
\begin{equation}
\bar{\mathcal{S}}_r(\vec{x})=\frac{1}{2}(\nabla\bar{\vec{u}}_r + \nabla\bar{\vec{u}}_r^T).
\label{eq:S-def}
\end{equation}
The remaining terms in the right-hand side are source and sink terms. In particular, the second term is the energy dissipation at scale $r$
\begin{equation}
\varepsilon_r(\vec{x}) \equiv \nu|\nabla\bar{\vec{u}}_r|^2;
\label{eq:dissipation}
\end{equation}
the third term is the conversion of potential energy to kinetic energy; and the last term corresponds to the exchange of energy due to external forces \citep[see][for details]{Aluie.2018}. These three terms  have different scale dominances and do not mix large scales with small scales. Indeed, energy dissipation dominates at small scales; external forcing is traditionally assumed to dominate at large scales\footnote{Notice, however, that the coupling at the ocean surface between the ocean and the atmosphere may span a wide range of scales in contrast with the standard assumptions in homogeneous three-dimensional turbulence.}; and the conversion between potential and kinetic energy, which is dominated by baroclinic instability and frontogenessis, may have some dominants scales, e.g. the Rossby radius of deformation. 

The focus of this study is at the range  between $\mathcal{O}(1\;\mathrm{km})$ and $\mathcal{O}(100\;\mathrm{km})$, i.e. at scales much larger than the dissipation scales, which implies that 
\begin{equation}
|\Pi_r(\vec{x})| \gg \varepsilon_r(\vec{x}).
\end{equation}
Although energy dissipation at the scale $r$ can be omitted in the equations at theses scales, dissipation at small scales is expected to control the flux of energy in the inertial range. 
Consequently, energy dissipation is at the core of theories explaining the energy cascade. In particular, according to Kolmogorov's Refined Similarity Hypothesis \citep{Kolmogorov.1962}, velocity differences 
\begin{equation}
\Delta_ru(\vec{x})\equiv | u(\vec{x}+ \vec{r}) - u(\vec{x})|
\label{eq:Du-definition}
\end{equation}
and coarse grained dissipation (equation \ref{eq:dissipation}) should be related by
\begin{equation}
\Delta_ru \sim (r\varepsilon_r)^\frac{1}{3},
\label{eq:refined-similarity-hypothesis}
\end{equation}
where $u(\vec{x})$ is an unspecified velocity component \citep{Frisch.1995, Davidson.2013}. It's worth mentioning that the symbol  $\sim$ in the above expression, and throughout the whole paper, omits any function of the position and, consequently, both sides of the expression may not have the same dimensions. 

\subsection{The multifractal hypothesis}

A fundamental property of turbulent flows is the non-vanishing of energy dissipation in the limit of zero viscosity (or infinite Reynolds number). 
This empirical results requires that velocity gradients must have singularities to keep a finite dissipation at infinite Reynolds numbers\citep{Eyink.2018}. 
This leads to the Multifractal Hypothesis of Turbulence \citep{Frisch.1995}.  According to it, in the limit of infinite Reynolds numbers, velocity gradients are characterized by local power-law scalings in the inertial range: at each point $\vec{x}$ there is an exponent  $h(\vec{x})\in [h_\infty, \infty)$, such that\footnote{Notice that $r$ must be also larger than the dissipation scale.} 
\begin{equation}
\overline{|\nabla u|}_r(\vec{x})\sim \left(\frac{r}{R}\right)^{h(\vec{x})},
\label{eq:h-definition}
\end{equation}
where $R$ is an integral scale and $r\ll R$. 
These exponents, also known as the H\"older exponents or Hurst exponents, are a measure of the regularity of the field.

The multifractal hypothesis for velocity gradients implies that the scaling exponents of velocity differences (equation \ref{eq:Du-definition}) are
\begin{equation}
\Delta_ru(\vec{x}) \sim \left(\frac{r}{R}\right)^{h(\vec{x}) + 1}
\label{eq:h-definition-Du}
\end{equation}
\citep[see][for a formal derivation]{Turiel.2008}. Moreover, armed with these scalings and using heuristic arguments, it is simple to derive the scaling for the coarse grained dissipation, i.e.
\begin{equation}
\varepsilon_r(\vec{x})  \sim \left(\frac{r}{R}\right)^{2h(\vec{x})},
 \label{eq:h-definition-diss}
\end{equation}
and the scaling for the local energy flux, i.e.
\begin{equation}
\Pi_r(\vec{x})\sim \left(\frac{r}{R}\right)^{3h(\vec{x})+2},
\end{equation}
\citep[see][for a rigorous derivation]{Eyink.1995, Eyink.2019}. 

\subsection{The singularity spectrum}

The singularity spectrum $D(h)$ is a function that for each value of the singularity exponent $h$, it provides the fractal dimension of the associated fractal component ${\cal F}_h$ (${\cal F}_h$ is the set of points that have the same singularity exponent $h$). This function determines the statistical properties of the flow.  Indeed, the probability of finding a given $h$  at the resolution $r$  is   
\begin{equation}
P_r(h)\propto \left(\frac{r}{R}\right)^{d-D(h)}.
\label{eq:pdf-h}
\end{equation}
Then, using the multifractal hypothesis (equation \ref{eq:h-definition}) and such probability, it is possible to estimate the behavior of the moments of the velocity gradients for $r\ll R$ as
\begin{equation}
\langle\overline{|\nabla u|}_r^p\rangle \sim\int_{\mathbb{R}^d}d \mu(h) \left(\frac{r}{R}\right)^{ph}\left(\frac{r}{R}\right)^{d-D(h)},
\end{equation}
where $d\mu(h)$ is a measure that gives the weight of the different exponents \citep{Frisch.1995}. The moments of velocity gradients have the property of self-similarity 
\begin{equation}
\langle\overline{|\nabla u|}_r^p\rangle \sim \left(\frac{r}{R}\right)^{\tau(p)},
\label{sup-eq:epsilon-mom}
\end{equation}
whose scaling function $\tau(p)$ is related to the $D(h)$ by a Legendre transform pair
\begin{equation}
\tau(p)=\min_{h}[ph+d-D(h)]
\label{sup-eq:Taup-Legendre-DSP}
\end{equation}
and
\begin{equation}
D(h)=\min_{p}[ph+d-\tau(p)],
\label{sup-eq:Dh-Legendre-DSP}
\end{equation}
as shown by \citet{Parisi.1985}. This transform pair implies that the singularity exponent  $h$ and the order of the moments $p$ are the derivatives of $\tau(p)$ and  $D(h)$ respectively
\begin{equation}
p=\frac{dD}{dh}
\label{eq:p-Dh}
\end{equation}
and
\begin{equation}
h=\frac{d\tau}{dp}
\label{eq:h-Tp}
\end{equation}
\citep[e.g.][]{Zia.2009}.

\begin{figure}[b]
\centering
 \noindent\includegraphics[width=19pc,angle=0]{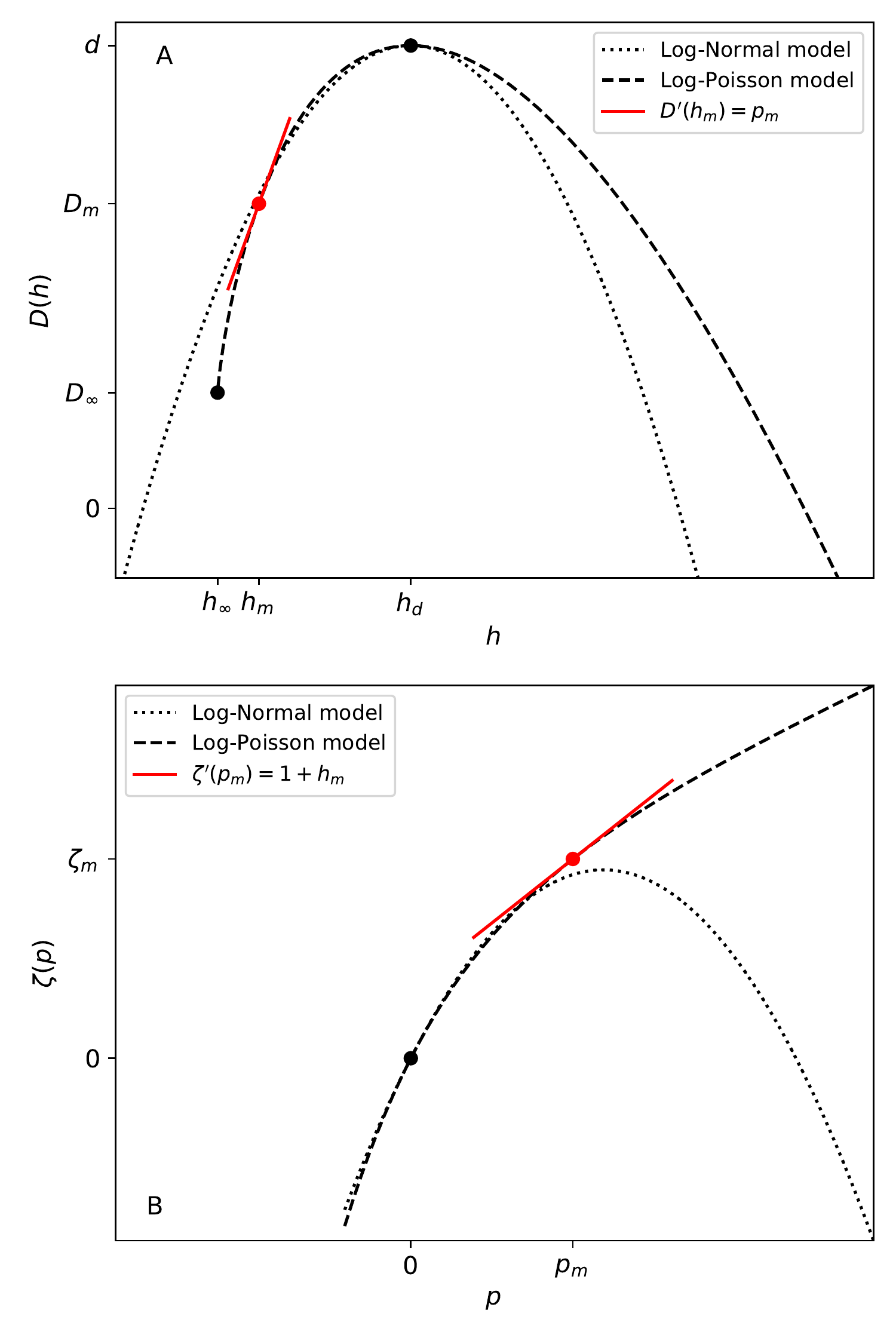}
\caption{(A) Singularity spectra corresponding to the Log-Normal and Log-Poisson models. (B) Scaling of the velocity structure functions predicted by Log-Normal and Log-Poisson models of velocity gradients (see equation \ref{eq:zetap-taup}).}
\label{fig:singspec_models}
\end{figure}

Moreover, one consequence of this scaling is that the moments of velocity differences, known as structure functions are also self-similar
\begin{equation}
S_p(r)\equiv\langle(\Delta_ru)^p\rangle\sim  \left(\frac{r}{R}\right)^{\zeta(p)}, 
\label{eq:zeta-definition}
\end{equation}
where $\zeta(p)$ is a continuous function of the moment order $p$. Indeed, using that
\begin{equation}
\langle(\Delta_ru)^p\rangle\sim  \langle\left( \frac{r}{R}\overline{|\nabla u|}_r\right)^p\rangle\sim \left(\frac{r}{R}\right)^{p +\tau(p)}
\end{equation}
and equation \ref{eq:zeta-definition}, the scaling of the velocity structure functions are given by
\begin{equation}
\zeta(p) = p + \tau(p)
\label{eq:zetap-taup}
\end{equation}
and, thus, they are also  determined by $D(h)$.
The singularity spectrum will also determine the behavior of any variable related to velocity gradients by a linear scaling when $r\ll R$ (see the appendix).


Within this framework, it is usually assumed that the mean value of $\overline{|\nabla u|}_r$  does not depend on the scale. This implies that 
\begin{equation}
\tau(1) \equiv 0,
\label{sup-eq:Tau1}
\end{equation}
which becomes a constraint over the singularity exponents. Indeed, using  equation \ref{sup-eq:Taup-Legendre-DSP}, it implies that 
\begin{equation}
d - D(h_1) = - h_1,
\label{eq:TransInvSI}
\end{equation}
where $h_1$ is the singularity exponent associated to the moment of order 1, so it verifies $D'(h_1) = 1$  \citep[e.g.][]{Turiel.2000}.

\begin{figure*}
\centering
 \noindent\includegraphics[width=39pc,angle=0]{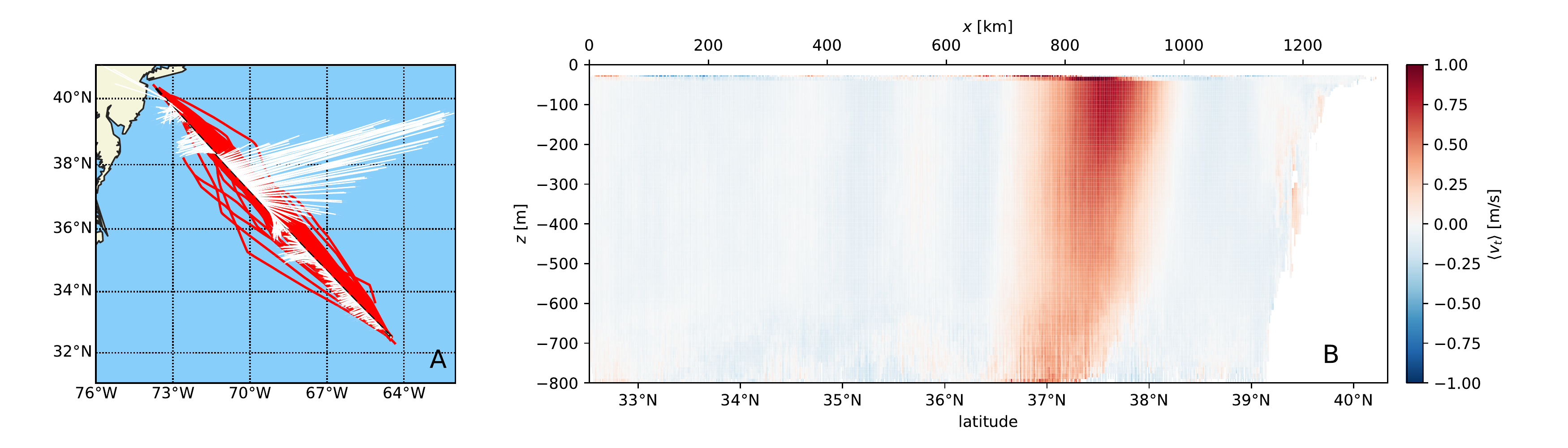}
\caption{Left. Oleander's courses with the mean velocity at 50 m depth superimposed. The dashed line indicates the reference course. Center. Available number of data at each depth. Right. Mean transverse velocity with the positive direction to the right of the reference track.}
\label{fig:data}
\end{figure*}

\subsection{Singularity spectrum models}

The characteristics of the energy cascade determine the functional dependency of $D(h)$ 
and, thus, the geometrical properties of the flow (fractal dimensions). As a consequence, different cascade models give rise to different singularity spectra that can be then compared to observations to get insight about the energy cascade \citep[see][for a review]{Seuront.2005}.  These models are constrained by the non-vanishing of energy dissipation at infinite Reynolds numbers. Indeed, according to  Onsager's theory of 1949, it implies that velocity differences (equation \ref{eq:h-definition-Du}) should have at least some singularities such that $h+1\leq1/3$ \citep{Eyink.2006, Eyink.2018}. Therefore, the singularity spectrum models should have singularity exponents equal or less than
\begin{equation}
h_0 = -\frac{2}{3}
\end{equation}
somehwere in the flow.

The simplest model that includes this constrain comes from Kolmogov's theory of 1941 \citep{Frisch.1995}, which corresponds to the case in which there is a single scaling exponent $h_0$, 
\begin{equation}
\tau(p) = h_0p
\label{eq:K41-Taup}
\end{equation}
and
\begin{equation}
D(h)= 
\begin{cases}
d & h=h_0\\
0 & h\neq h_0
\end{cases}.
\label{eq:K41-Dh}
\end{equation}
Then, the scaling of the structure functions is linear $\zeta(p)=(h_0+1)p$ with the slope of 1/3 predicted by Kolmogorov. Nevertheless, the observed velocity structure functions deviate from this linear scaling giving rise to the so-called anomalous scaling. As a consequence, turbulent regimes cannot have a single fractal component, i.e. a unique singularity exponent, but they must have a range of exponents implying that they  are mutifractal. 

It is important to recall that not only isotropic three-dimensional turbulence, but also geophysical turbulence is multifractal. Indeed, Onsager argument implies the existence of singularities in the flow and the observed anomalous scaling of the oceanic velocity structure functions at scales larger than the Ozmidov length \citep[see figure 5 in][]{Sukhatme.2020} imply that a range of singularities must be present. In addition to this argument, its worth mentioning that in previous studies it has already been stablished the validity of the multifractal framework in the ocean for scales covering from the largest submesoscales to the large scales \citep[see figure 2 in][]{Isern.2007}. Nevertheless, the singularity spectrum model that characterizes ocean turbulence in the submesoscale to mesoscale range still has to be determined. 


A starting point to generate a multifractal turbulent cascade is to describe it as a multiplicative process of independent random variable identically distributed \citep{Frisch.1995}. Different  probability density function used to characterize their variability generate different $\tau(p)$ and $D(h)$. Here, we focus on the two most extended models: the Log-Normal model \citep{Kolmogorov.1962}, which is still widely used in geophysical and three-dimensional turbulence \citep[e.g.][]{Pearson.2018, Dubrulle.2019}; and the Log-Poisson model \citep{She.1995}. Mathematically, the Log-Normal model assumes that the logarithm of the random variables have a Normal distribution while for the Log-Poisson model they follow a Poisson distribution. Although we do not present it here, it is worth mentioning the log-L\'evy model \citep{Schertzer.1987}, which is a generalization of the log-Normal model.

The Log-Normal model (figure \ref{fig:singspec_models}) is characterized by two parameters: the mode $h_d$ and the standard deviation $\sigma_h$ given by
\begin{equation}
\tau(p) = h_d p - \frac{1}{2}\sigma_h^2p^2
\label{sup-eq:Log-Normal-Taup}
\end{equation}
and
\begin{equation}
D(h)=d-\frac{1}{2}\left(\frac{h-h_d}{\sigma_h}\right)^2.
\label{sup-eq:Log-Normal-Dh}
\end{equation}
The singularity spectra derived from the Log-Normal model is symmetric about the mode, as it is evident from equation \ref{sup-eq:Log-Normal-Dh}.

The Log-Poisson model (figure \ref{fig:singspec_models}) is characterized by three parameters: the most singular exponent $h_\infty$, its fractal dimension $D_\infty$ and the intermittency parameter $\beta$,
\begin{equation}
\tau(p) = h_\infty p + (d-D_\infty)(1-\beta^p) 
\label{eq:Log-Poisson-Taup}
\end{equation}
from where:
\begin{equation}
D(h) = D_\infty - \frac{h-h_\infty}{\ln \beta}\left[1 - 
\ln\left(-\frac{h-h_\infty}{(d-D_\infty)\ln\beta} \right)\right].
\label{eq:Log-Poisson-Dh}
\end{equation}
The condition given by equation \ref{eq:TransInvSI} imposes an additional constraint:
\begin{equation}
\beta = 1 + \frac{h_\infty}{(d - D_\infty)},
\label{eq:Log-Poisson-beta}
\end{equation}
which allows to reduce the parameters of this model to only two. An important feature of the Log-Poisson model is that their parameters have a physical interpretation \citep{She.1995}, rather than being \textit{ad hoc} parameters to be fitted. According to this model,  $\beta$ measures the fraction of energy retained along the cascade \citep{She.1994,She.1995,Turiel.2000}. 
This model, contrary to the Log-Normal model, has a singularity spectrum that is asymmetric about the mode.

\section{Data}
\label{sec:data}

\subsection{ADCP measurements}

\begin{figure}
\centering
 \noindent\includegraphics[width=19pc,angle=0]{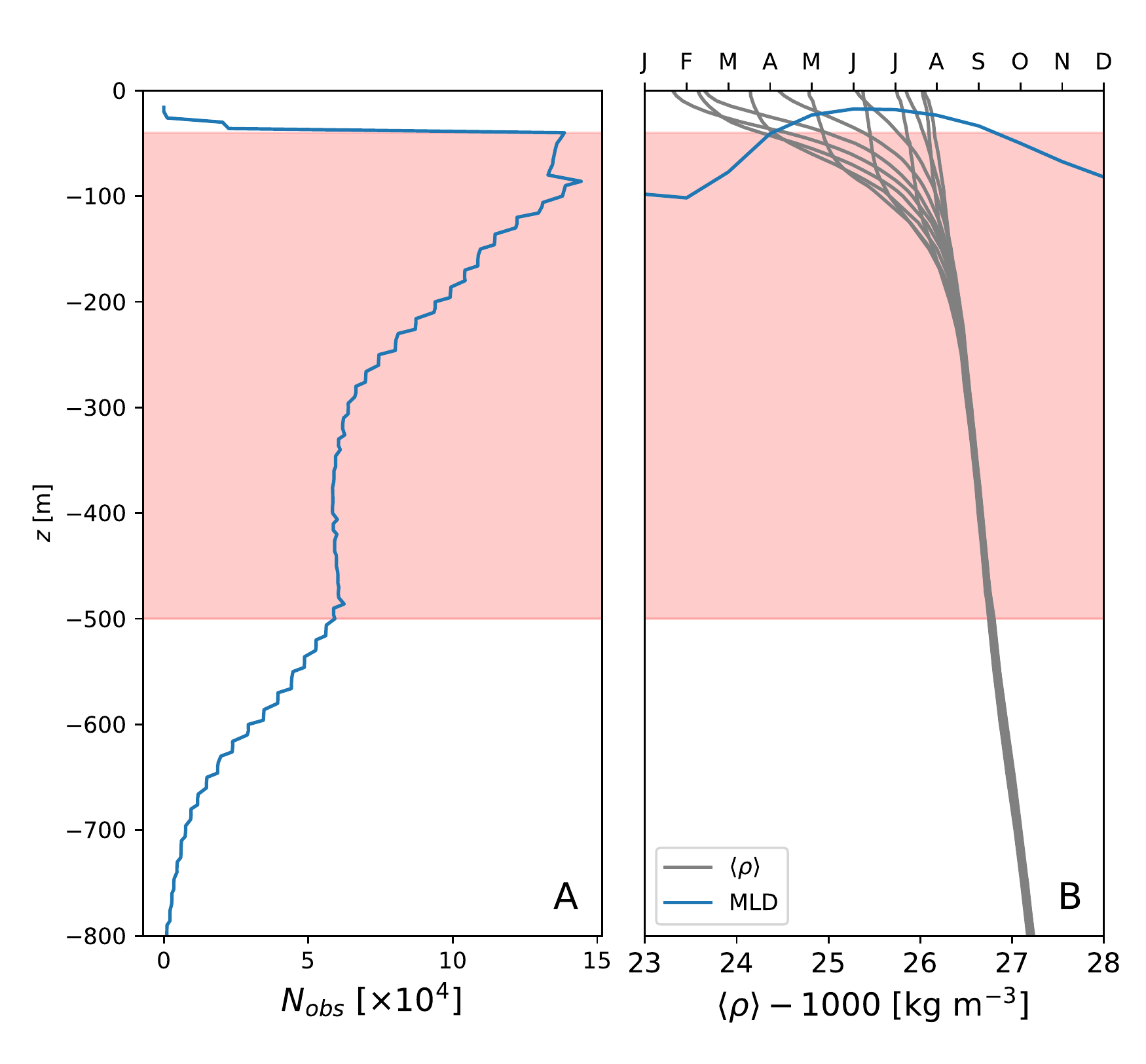}
\caption{(A) Number of observations at each depth. (B) Monthly mean vertical profiles of density derived from the WOA18 climatology using the TEOS-10 Equation of State for Sea Water (gray) and temporal evolution of the mean MLD (blue). The red band corresponds to the range of depths analyzed in this study.}
\label{fig:available_data}
\label{fig:seasonal_mld}
\end{figure}

In this study we analyzed 11 years (2005-2016) of measurements provided by the Oleander project \citep{Rossby.2010}, which consists on 360 weekly profiles at approximately 2 km resolution of horizontal velocities between New Jersey and Bermuda in the North Atlantic across the Gulf Stream  measured by a ship-of-opportunity (see Figure \ref{fig:data}).  The original velocity measurements can be downloaded at http://po.msrc.sunysb.edu/Oleander/. Data prior to 2005 were discarded because they correspond to a different instrument with lower performance. Velocities were decomposed into a longitudinal $u_l(x, z)$, along ship track;  and  transverse $u_t(x, z)$, perpendicular to ship track component (with positive values to the right). These measurements covered a range of depths between 40 m and 800 m. However, due to the scarcity of data below 500 m (figure \ref{fig:available_data}A), the study was restricted to the range of depths between 40 and 500 m.  

\subsection{Ancillary data}

Climatological values of vertical density variations were calculated from  the World Ocean Atlas 2018 at 1/4$^\circ$ of resolution produced by NOAA (https://www.nodc.noaa.gov/OC5/woa18/).  We used monthly salinity and temperature profiles to first generate Absolute Salinity and Conservative Temperature and, then, compute the  potential density anomaly with reference pressure of 0 dbar using the Python-GSW package (https://github.com/TEOS-10/python-gsw/). Resulting density fields were horizontally averaged over the box [31$^\circ$N, 39$^\circ$N]$\times$[76$^\circ$W, 62$^\circ$W] to produce the vertical profiles shown in Figure \ref{fig:seasonal_mld}B. 

This data set was complemented with the Mixed Layer Depth (MLD) climatology produced by Ifremer (http://www.ifremer.fr/cerweb/deboyer/mld/home.php). Original data was also averaged over the same box as the WOA climatology to obtain the mean MLD shown in Figure \ref{fig:seasonal_mld}B.

\begin{figure*}
\centering
 \noindent\includegraphics[width=39pc,angle=0]{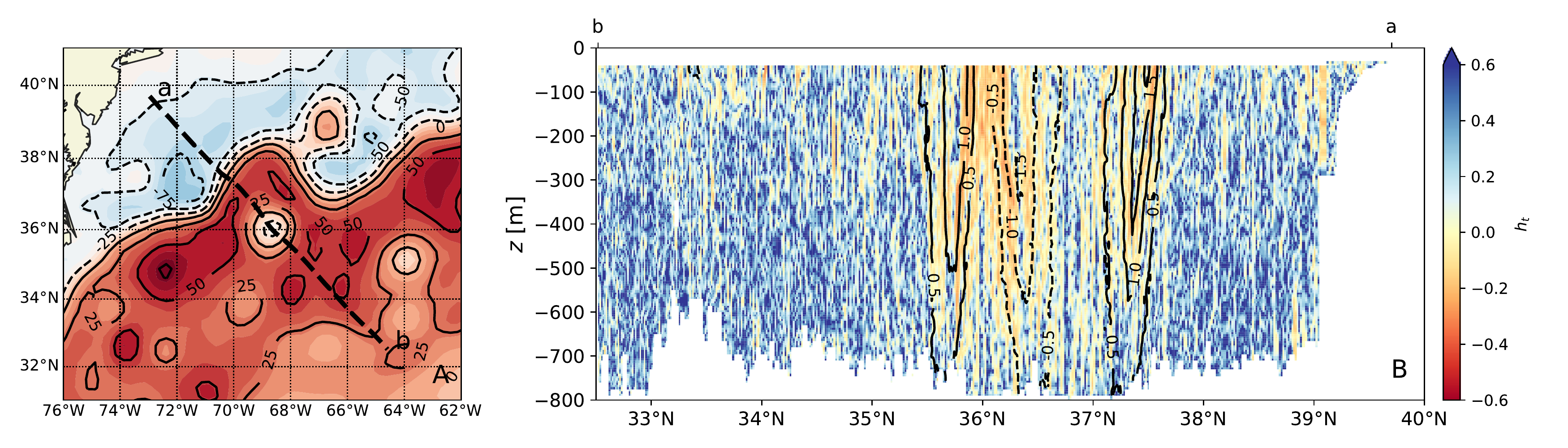}
\caption{(A) Sea level map obtained from the measurements of multiple altimeters for May 3 2008. The thick dotted line corresponds to the ship's track. (B) Singularity exponents derived from transverse velocity around the date of the sea level field. White areas correspond to missing data. Black lines correspond to the transverse velocity in m/s (positive Northeastwards). }
\label{fig:example-exps}
\end{figure*}

Additional altimetric data was used to help interpret ADCP data (see the example in Figure \ref{fig:example-exps}). In particular, Sea Surface Heights were obtained from SSALTO/DUACS Delayed-Time Level-4 data measured by multi-satellite altimetry observations over Global Ocean. Data are available at http://marine.copernicus.eu.

\section{Methods}
\label{sec:methods}

\subsection{Retrieval of singularity exponents}

Singularity exponents were computed from velocity gradients using equations \ref{eq:coarse-grained} and \ref{eq:h-definition},  where $G(\vec{x})$ is the mother wavelet, which does not have the requirement of the admissibility condition \citep{Turiel.2000}, and $d=2$.  Then, from the coarse grained field at the resolution of observation $r_0$, which is the smallest accessible scale, the singularity exponents were obtained as
\begin{equation}
h(\vec{x})\approx \frac{\log \overline{|\nabla u|}_{r_0} - \log \langle\overline{|\nabla u|}_{r_0}\rangle}{ \log (r_0/R)}
\label{eq:sings-algorithm}
\end{equation}
(see \citet{Turiel.2008}).

In this study, we used the algorithm described in  \citet{Pont.2013} to compute singularity exponents with two modifications. First, the anisotropy between horizontal and vertical scales was taken into account by computing gradients in physical coordinates rather than pixel coordinates. Second, the smallest accessible scale $r_0$ was computed as 
\begin{equation}
\frac{r_0}{R} \approx \frac{1}{\sqrt[d]{N_{obs}}},
\label{eq:smallest-effective-scale}
\end{equation}
where $N_{obs}$ is the number of valid observations in a single transect. It is worth mentioning that, the existence of gaps in real data increases the smallest accessible resolution. 

\subsection{Estimation of singularity spectra}

Approximating the probability of finding an exponent $h$ at the resolution of observation $r_0$ (equation \ref{eq:pdf-h}) by its histogram, the singularity spectrum associated to a given sample of data can be estimated as
\begin{equation}
D(h_i) \approx d -  \frac{\log N_i -\log N_{max}}{\log (r_0/R)}, 
\label{eq:D(h)-algorithm}
\end{equation}
where $N_i$ is the number of observations within the bin $[h_i-\frac{\delta h}{2}, h_i+\frac{\delta h}{2})$ and $N_{max}=\mbox{max}_i \{N_i\}$. As the support of the gradient is the whole domain, the largest fractal dimension is the dimension of the whole space $d$, 
\begin{equation}
D(h_d)=\max(D(h))\equiv d, 
\end{equation}
where $h_d$ is the mode of the  singularity exponents. Moreover,  the error bars  of the singularity spectrum were estimated as
\begin{equation}
\delta D_i = \frac{3}{|\log (r_0/R)| N_i^{\frac{1}{2}}}
\label{eq:error-D(h)}
\end{equation}
\citep[see][]{Turiel.2006}.

\subsection{Singularity analysis of ADCP data}

Vertical profiles of singularity exponents for the longitudinal and transverse components of velocity, $h_l(x, z)$ and $h_t(x, z)$ respectively, were computed for each velocity transect. Besides,  in the limit for $r\ll R$, the coarse grained dissipation will be dominated by the most singular exponent and, thus,  
\begin{equation}
h(x, z)=\min\left(h_l(x, z),h_t(x, z)\right)
\end{equation}
Nevertheless, we have retained throughout the paper the values of the singularity exponents of both velocity components\footnote{In this paper, two classes of subindex are used for singularity exponents. On one side, exponents $h_0$, $h_1$, $d$ and $h_\infty$ refer to specific values. On the other side, $h_l$ and $h_t$ refer to the set of components of the longitudinal and traverse components. If necessary they can be combined, i.e. $h_{t\infty}$ is the most singular exponent of the transverse velocity component.}. Singularity spectra  were obtained from the associated singularity exponents south of  the continental shelf, which is located at 39$^\circ$N (figure \ref{fig:example-exps}B), using equation \ref{eq:D(h)-algorithm}.  Singularity exponents at the borders of the domain or a data gap were removed, as the estimate of its value is of lower quality. In order to impose the translational invariance and correct any shift that may exist in the singularity exponents, two-dimensional singularity spectra were calculated for each transect ($d=2$ and $\delta h=0.05$) and equation \ref{eq:TransInvSI} was used.  Then, one-dimensional singularity spectra ($d=1$, $\delta h=0.1$) were computed for each transect and at each depth in order to explore the vertical variations of intermittency. 

\section{Analysis of the empirical singularity spectra}
\label{sec:obs}

\subsection{An example}

Figure \ref{fig:example-exps} shows an example of a vertical profile of transverse singularity exponents $h_t(x,z)$. Sea Surface Height shown in figure \ref{fig:example-exps}A provide a view of the flow topology simultaneous to ship measurements. In this example, the ship crossed the Gulf Stream (between 37$^\circ$N and 38$^\circ$N) and a mesoscale vortex centered at 36$^\circ$N. Singularity exponents (figure \ref{fig:example-exps}B) exhibit a strong vertical coherence of those singularities associated to the Gulf Stream and those related to the vortex, which are located within the maxima of velocities. 

The scaling of the the local fluxes of energy implies that at different scales they are related by
\begin{equation}
\Pi_{r_2}(\vec{x})\approx \left(\frac{r_2}{r_1}\right)^{3h(\vec{x})+2}\Pi_{r_1}(\vec{x}),
\label{eq:scaling-Pi}
\end{equation}
with $r_2<r_1$. 
Based on the above relation, the spatial distribution of singularity exponents (for instance, those in figure \ref{fig:example-exps}) maps the pathways through which energy is transferred towards smaller scales. Contrary to the computation of the local energy flux $\Pi_r(\vec{x})$, it is not necessary to know the velocity at small scales and can be used to analyze real observations \citep[see][and the discussion there]{Aluie.2018}. For sure, singularity exponents cannot quantify the flux of energy nor even estimate its sign. Nevertheless, at the resolution of the data ($\sim 2$ km???) we expect a direct energy cascade \citep{McWilliams.2016}. At coarser resolutions ($> 10$ km) there is an inverse cascade and singularity exponents may not be the same. To test for this change, we would need more data than a velocity transects to keep the accuracy and precision of $h$ reasonable \citep{Pont.2013}.

\begin{figure}
\centering
 \noindent\includegraphics[width=19pc,angle=0]{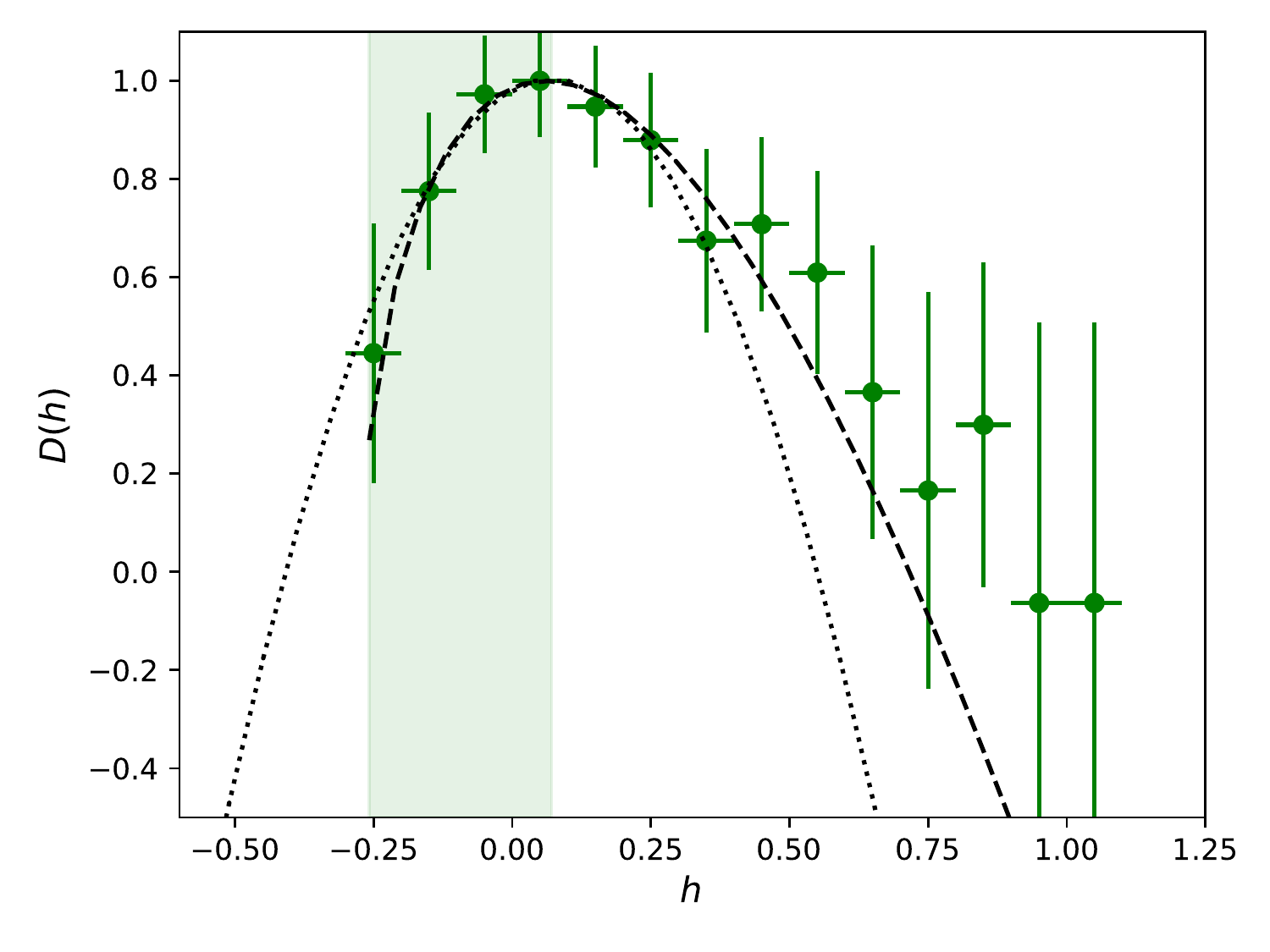}
\caption{Singularity spectrum at 100 m depth (dots) for the same transect with the corresponding error estimations (vertical lines) and bin width (horizontal lines). The light green interval corresponds to $\Delta h^-$, bounded by $h_\infty$ to the left and by $h_d$ to the right. Dashed line corresponds to the Log-Poisson model obtained using the observed $h_\infty$ and the $\beta$ obtained at that depth. The dotted line corresponds to a parabola fitted to the data.}
\label{fig:example-spectra}
\end{figure}

The singularity spectrum  at 100 m (figure \ref{fig:example-spectra}) shows a characteristic convex curve, from which two features are evident: the asymmetry of this singularity spectrum and the existence of a minimum singularity $h_\infty$ with a fractal dimension $D(h_\infty)  > \min(D(h))$. Recall that, we are analyzing a one-dimensional transect of a three-dimensional flow, implying that negative dimensions have to be interpreted as occurrence probabilities \citep{Mandelbrot.1990}.  Moreover, the singularity spectrum does not change with the change of the spatial resolution, although as resolution is getting coarser the statistics is reduced and therefore the resulting spectra get noisier and the error bars larger, which key characteristic of multifractal systems. In \citet{Turiel.2008} you can find an extensive discussion on the validity of the multifractal approach in many different contexts.

\subsection{Characterization of singularity spectra}

The singularity spectrum shown in the example of figure \ref{fig:example-spectra} is representative of the spectra observed at different depths and times. A systematic characterization of each singularity spectrum, i.e. for each depth and transect, has been done using four shape parameters: 
\begin{enumerate}

\item Pearson's first skewness coefficient or mode skewness $s_d$, i.e. the skewness about the mode $h_d$,
\begin{equation}
s_d(z, t)\equiv \frac{\langle h\rangle - h_d}{\sigma_h},
\label{eq:mode-skewness}
\end{equation}
where $\langle h\rangle$ is the mean value and $\sigma_h$ the standard deviation. The variable $z$ refers to the depth and $t$ to the central time of the transect from where the singularity exponents $h$ were calculated; 

\item the most singular exponent $h_\infty$,
\begin{equation}
h_\infty(z, t) \equiv \min(h);
\label{eq:most-singular-def}
\end{equation}

\item the fractal dimension of its set $D_\infty$,
\begin{equation}
D_\infty(z, t) \equiv D(h_\infty);
\label{eq:most-singular-dim-def}
\end{equation}

\item the width of the singularity spectrum $\Delta h^-$, defined as
\begin{equation}
\Delta h^-(z, t)\equiv h_d - h_\infty,
\label{eq:anomalous-def}
\end{equation}
which is a measure of the anomalous scaling. 

\end{enumerate}

\begin{figure}[b]
\centering
 \noindent\includegraphics[width=19pc,angle=0]{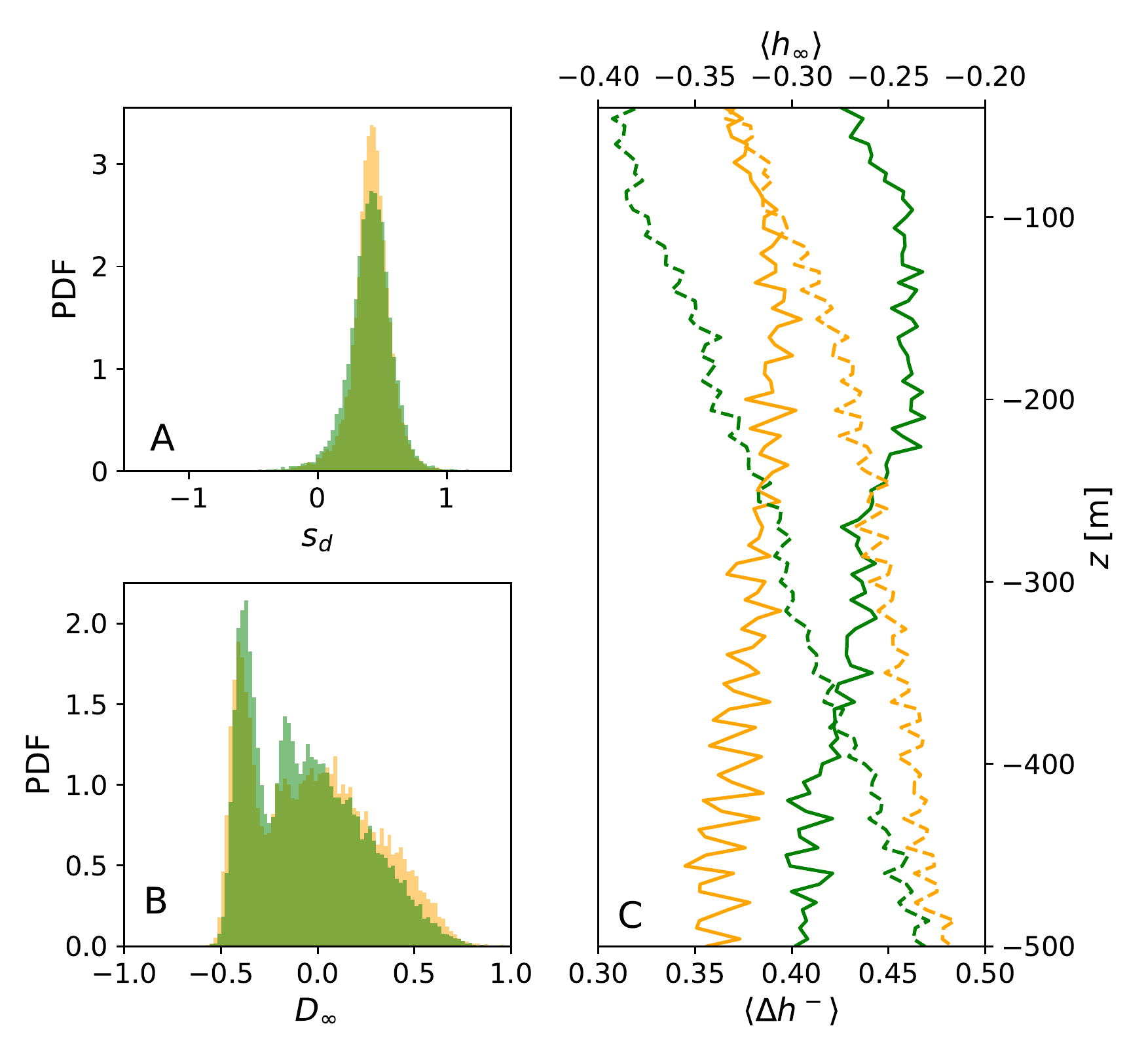}
\caption{(A) observed distributions of mode skewness $s_d$ and (B) fractal dimension of the most singular set $D_\infty$. (C) time-averaged vertical variation of anomalous scaling $\langle \Delta h^-\rangle$ (solid lines) and most singular exponent $\langle h_\infty\rangle$ (dashed lines). Green and orange colors correspond to transverse ($h_t$) and longitudinal ($h_l$) singularity exponents, respectively.}
\label{fig:singspec_properties}
\end{figure}

Indeed, the value of singularity exponent $h_d$ associated to the maximum of $D(h)$ corresponds  to $p=0$ (horizontal tangent line to $D(h)$) and $h_\infty$ corresponds to the largest (in absolute value) negative slope of $\tau(p)$ and thus to the largest moments $p\rightarrow \infty$. Recall that $D(h)$ is a convex function and $p$ and $h$ are the slopes of $D(h)$ and $\tau(p)$, respectively (equations \ref{eq:p-Dh} and \ref{eq:h-Tp}). Consequently, equation \ref{eq:anomalous-def} can be rewritten using  equations \ref{eq:h-Tp} and \ref{eq:zetap-taup} as
\begin{align}
\Delta h^- &= \left.\frac{d\tau}{dp}\right|_{p=0} -\left. \frac{d\tau}{dp}\right|_{p\rightarrow\infty}\\ 
                 & = \left.\frac{d\zeta}{dp}\right|_{p=0} - \left.\frac{d\zeta}{dp}\right|_{p\rightarrow\infty} \label{eq:Dh-to-anomalous}
\end{align}
that it, $\Delta h^-$ is a measure of the difference between the slopes of $\zeta(p)$ at the origin and at the largest orders of the structure function. In our calculations, the mode $h_d$ was estimated by adjusting a parabola in a small region around the maximum of $D(h)$, and then calculating its maximum from the adjusted curve.

\begin{figure}[b]
\centering
 \noindent\includegraphics[width=19pc,angle=0]{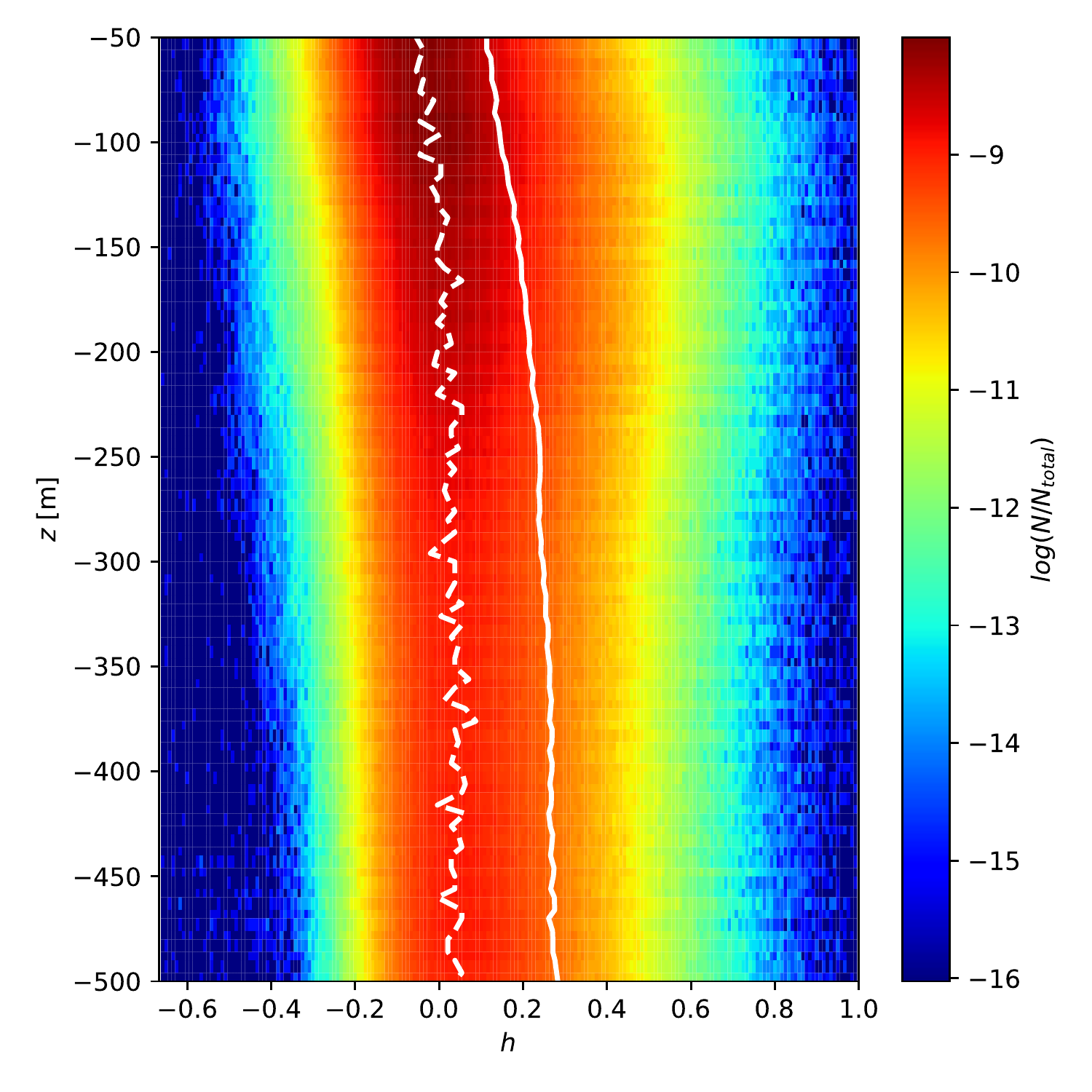}
\caption{Empirical histograms of singularity exponents at each depth. The range of values for the singularity exponents go from the Kolmogorov exponent $h_0=-2/3$ to 1 in bins of 0.008. The histogram relative frequencies $N/N_{total}$ are computed during the whole period analyzed and are logarithmically represented with a color code. The solid line corresponds to the vertical variation of the mean singularity exponent while the dashed line corresponds to the vertical variation of the most probable value (mode).}
\label{fig:pdf2dh}
\end{figure}

\begin{figure*}
\centering
 \noindent\includegraphics[width=39pc,angle=0]{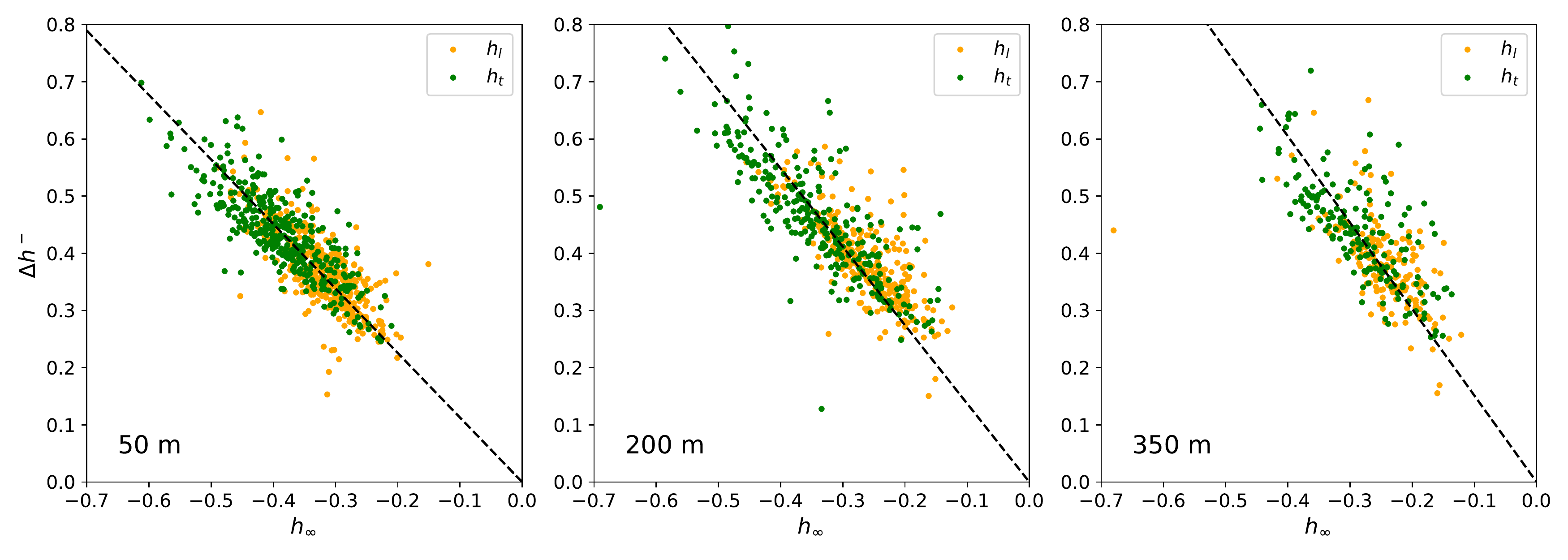}
\caption{Scatter plot between the anomalous scaling measure $\Delta h^-$ and the singular exponents $h_\infty$ at 50, 200 and 350 m, respectively. The dashed line corresponds to the fitted straight line. Green and orange colors correspond to transverse ($h_t$) and longitudinal ($h_l$) singularity exponents, respectively. }
\label{fig:anomalous_vs_hoo}
\end{figure*}

\subsection{Analysis of the shape parameters}

Mode skewness $s_d$ (equation \ref{eq:mode-skewness}) is a measure of the symmetry around the mode and can be used to test the symmetry conjecture \citep{Dubrulle.2019}. Figure \ref{fig:singspec_properties}A shows the histogram of $s_d$ at all depths for both, the transverse and the longitudinal velocity components, revealing that $s_d$ has a negligible probability of being negative or even zero, thus evidencing that the singularity spectra are almost always asymmetric.  

The most singular exponent $h_{\infty}$ (equation \ref{eq:most-singular-def}) measures the strength of the most intense velocity gradient. Its value ranges from -0.6 to -0.1, with a tendency to decrease with depth; transverse and longitudinal components have different $h_{\infty}$, with the first being more singular (smaller $h$, figure \ref{fig:singspec_properties}C).  Moreover, the histograms for $h_\infty$ are dominated by the contribution of the Gulf Stream and the mesoscale vortices detached from it, as it can be seen in the example of figure \ref{fig:example-exps}B. 

The fractal dimension of the most singular set $D_{\infty}$ (equation \ref{eq:most-singular-dim-def})  spreads between -0.5 and 0.25 (figure \ref{fig:singspec_properties}B), which would correspond to values between 1.5 and 2.25, if the singularity spectrum would have been computed in three dimensions. The distribution of $D_\infty$ shows a peak around $D_\infty\approx -0.4$ due to those sets composed of a single point, which is indicative of the need to have more observations to accurately estimate this dimension (see equation \ref{eq:error-D(h)} and the error bars in figure \ref{fig:example-exps}C). Besides, dissipative structures defined by $h_\infty$  are not filamentary as in \citet{She.1994}, i.e. have fractal dimensions larger than one ($D_\infty > -1$ for $d=1$, figure \ref{fig:singspec_properties}B). 

\begin{figure}[b]
\centering
 \noindent\includegraphics[width=19pc,angle=0]{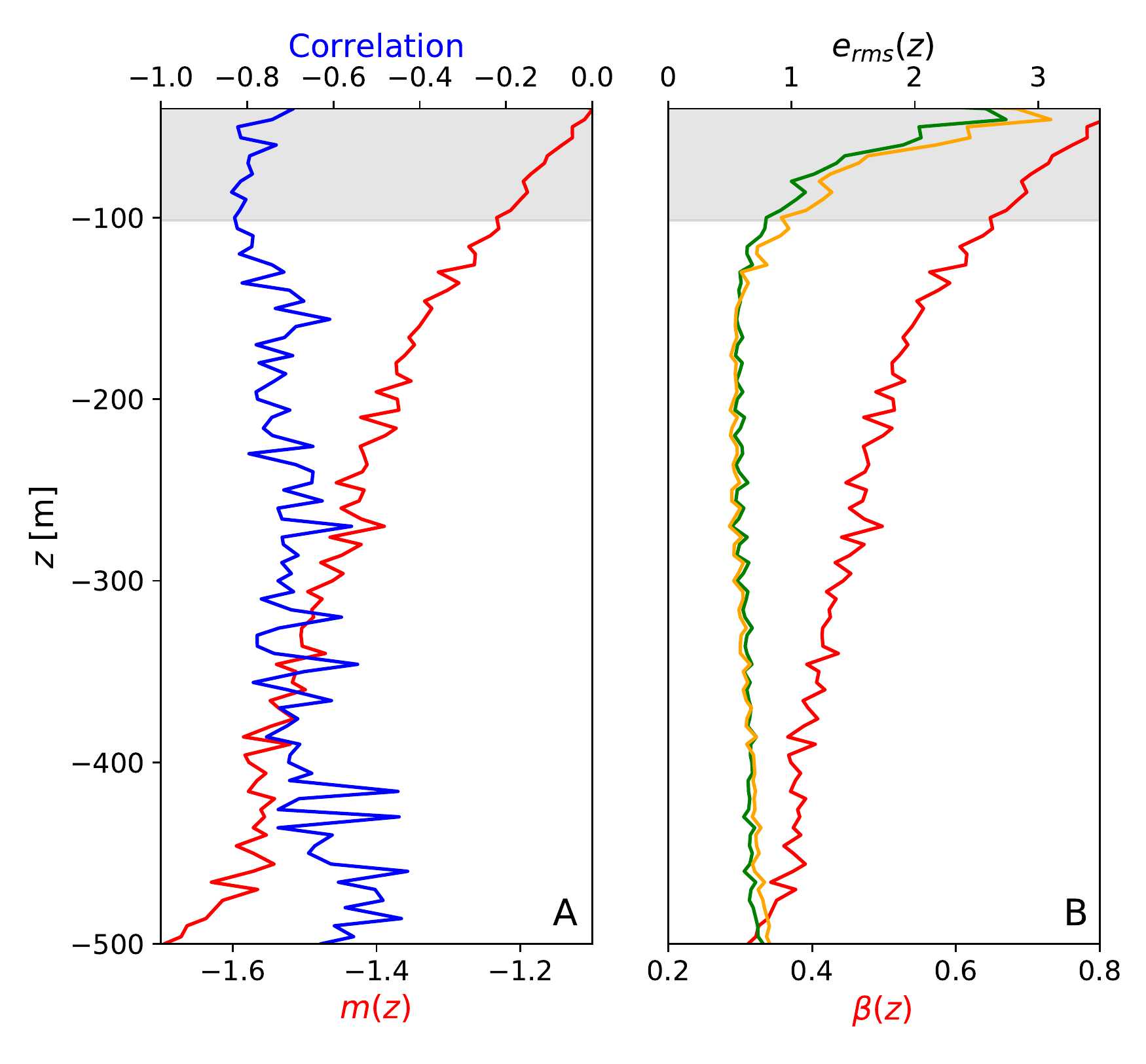}
\caption{ (A) Slope $m(z)$ (red) and linear correlation parameter between $h_\infty$  and $\Delta h^-$ (blue). (B) Intermittency parameter, $\beta$, (red) and normalized RMSE (orange and green). Green and orange colors correspond to traverse ($h_t$) and longitudinal ($h_l$) singularity exponents, respectively. The gray band indicate the depths that can be within the Mixed Layer}
\label{fig:logpois_assesment}
\end{figure}

The width of singularity spectrum spreads from 0.2 to 0.7 approximately, suggesting that they have significant variability, with the spectra of the transverse velocity component  being wider (figure \ref{fig:anomalous_vs_hoo}). The time-averaged amplitude of the singularity spectrum $\langle \Delta h^-\rangle$ is maximum at depths between 100 m and 250 m  (figure \ref{fig:singspec_properties}C). Moreover, the comparison between all the observed $\Delta h^-$ and $h_\infty$ at each depth shows a decreasing linear dependence between them (figure \ref{fig:anomalous_vs_hoo}):
 \begin{equation}
\Delta h^-\approx m(z) h_\infty,
\label{eq:linear-model-obs}
\end{equation}
with $m(z) < 0$. We estimate the value of $m(z)$ at each depth $z$ by means of a least square regression of $\delta h^-(z)$ vs $h_{\infty}(z)$ obtained from both, longitudinal and traverse velocity components. Results are shown in figure \ref{fig:logpois_assesment}A, along the linear correlation parameter between $\Delta h^-$ and $h_\infty$. Largest correlations, of the order of -0.8, are found in the upper layers of the ocean, were the range of values for $h_\infty$ is larger, and reduce to values of the order of -0.75 below 150 m. In the lower layers, where the ranges of values of both $\Delta h^-$ and $h_\infty$ are smaller, the linear correlations are of the order of -0.6. 

On the other hand, the vertical variation of the shape parameters ($h_\infty$ and $\Delta h^-$ in figure \ref{fig:singspec_properties}) and the histogram of singularity exponents (figure \ref{fig:pdf2dh}) imply changes in the energy transferred between scales as it is evident from equation \ref{eq:scaling-Pi}. 
The Kolmogorov singularity exponent $h_0=-2/3$ (see section \ref{sec:theory}) would imply that that all energy is transferred through scales without loses. Nevertheless, as shown in figure \ref{fig:pdf2dh}, $h_0$ rarely takes the value $-2/3$ and when it does it is likely due to the errors in the estimation of $h$.

\section{Analytical model}
\label{sec:model}

Our results show that singularity spectra are asymmetric (figure \ref{fig:singspec_properties}A), implying that those cascade models generating symmetric distributions of singularity exponents should be discarded. This is the case of the Log-Normal model, widely applied to three-dimensional turbulence \citep{Dubrulle.2019}. The Log-Poisson model, on the contrary, is asymmetric and verifies $D_\infty > \min(D(h))$.  Figure \ref{fig:example-spectra}C shows an example of the better fitting of the Log-Poisson model to the observed singularity spectrum than the Log-Normal. Consequently, we used it to derive the analytical relationship between $h_\infty$ and  $\Delta h^{-}$ (equation  \ref{eq:linear-model-obs}).  

\subsection{Slope between $\Delta h^-$ and $h_\infty$}

Using that $D(h_d) \equiv d$, the Log-Poisson model (equation \ref{eq:Log-Poisson-Dh}) gives 
\begin{equation}
d = D_\infty - \frac{h_d-h_\infty}{\ln \beta}\left[1 - 
\ln\left(-\frac{h_d-h_\infty}{(d-D_\infty)\ln\beta} \right)\right].
\end{equation}
Then, introducing the definition of $\Delta h^-$ (equation \ref{eq:anomalous-def}) and reordering terms; we obtain
\begin{equation}
-\frac{(d - D_\infty)\ln\beta }{\Delta h^-} = 1 - 
\ln\left(-\frac{\Delta h ^-}{(d-D_\infty)\ln\beta} \right).
\end{equation}
The solution to this equation is
\begin{equation}
\Delta h^- = -(d - D_\infty)\ln\beta.
\end{equation}
and, inserting equation \ref{eq:Log-Poisson-beta},  it becomes
\begin{equation}
\Delta h^- = \frac{\ln \beta}{1-\beta}  h_\infty.
\label{eq:linear-model-log-poisson}
\end{equation}

Notice that the Log-Normal model does not provide any prediction for this relation because in this model $h_{\infty}=-\infty$.  Then, comparing equations  \ref{eq:linear-model-obs} and \ref{eq:linear-model-log-poisson} we get that 
\begin{equation}
m(z)= \frac{\ln \beta}{1-\beta}, 
\end{equation}
which reveals that  the vertical decrease of observed slopes results in a reduction of the intermittency parameter  (figure \ref{fig:logpois_assesment}B). As a consequence, the dependence of intermittency with depth (figure \ref{fig:logpois_assesment}A) is interpreted as  vertical changes of the processes underlying the cascade, whose competition with  changes in dissipation (figure \ref{fig:singspec_properties}C) leads to a maximum anomalous scaling of velocity structure functions between 100 and 250 m (figure \ref{fig:singspec_properties}C). 

The relationship between anomalous scaling and dissipation can be, then,  retrieved from equation \ref{eq:linear-model-log-poisson} (or equivalently  equation \ref{eq:linear-model-obs}) introducing the relation between the amplitude of the singularity spectrum and the change of slope in the structure functions scaling  (equation \ref{eq:Dh-to-anomalous}), on one side; and singularity exponents of dissipation 
\begin{equation}
h_\varepsilon(\vec{x}) \equiv 2h(\vec{x}),
\label{eq:h-to-diss}
\end{equation}
on the other side;
\begin{equation}
\left. \frac{d\zeta}{dp}\right|_{p=0} - \left. \frac{d\zeta}{dp}\right|_{p\rightarrow\infty} =  \frac{1}{2}\frac{\ln\beta}{1-\beta}\min(h_\epsilon),
\label{eq:linear-model-Struct-Disp}
\end{equation}
which states that anomalous scaling is a function of the intermittency parameter and the most singular exponent of dissipation. 
 Notice that the above relation is independent from Kolmogorov's Refined Similarity Hypothesis (equation \ref{eq:refined-similarity-hypothesis}). Although, it could be rewritten using to it, or for any variable whose singularity exponents are linearly related to $h(\vec{x})$ (see the appendix).

\subsection{Model assessment}

To get further evidence on the validity of the Log-Poisson model, the normalized Root Mean Square Error  (RMSE) has been defined as the RMSE of the differences between it (equation \ref{eq:Log-Poisson-Dh}) and observations (equation \ref{eq:D(h)-algorithm}) divided by the standard deviation of observational error estimates (equation \ref{eq:error-D(h)})
\begin{equation}
e_{rms} =\left[ \frac{\sum_{i} [D_{obs}(h_i)-D_{theory}(h_i)]^2}{\sum_{i}\delta D_i^2}\right]^\frac{1}{2}.
\end{equation}

We have found that it is smaller than one only for depths below 100 m (figure \ref{fig:logpois_assesment}B), implying that above this depth the deviations of our fit of the Log-Poisson model from observations does not fall within the estimated errors. The existence of such discrepancies may be explained by the presence of the Mixed Layer (ML), only partly captured by Oleander data. Indeed, observations are within the ML only during the period October-March approximately for depths shallower than 100 m (figure \ref{fig:seasonal_mld}B). The lack of an instantaneous determination of the ML prevents to separately investigate the energy cascade within it. Besides, the maximum amplitude of the singularity spectra i.e., the strongest anomalous scaling, corresponds to depths dominated by the seasonal variation of the pycnocline (40 - 200 m, figure \ref{fig:seasonal_mld}B).  It is worth mentioning, however, that even at these depths singularity spectra are still asymmetric and $h_\infty$ and $\Delta h^-$ tend to be linearly related (see figure \ref{fig:logpois_assesment}A).

\section{Discussion}
\label{sec:discussion}

The multifractal theory of turbulence outlined in the previous sections is based on three main contributions: Onsager's theory of ideal turbulence, as presented by \citet{Eyink.2018}; the multifractal formalism described in \citet{Frisch.1995}; and different models of turbulent cascade, mainly from \citet{Kolmogorov.1962} and \citet{She.1994}. Nevertheless, we have chosen to built it from the singularity exponents of velocity gradients $\overline{|\nabla u|}$ instead of velocity differences $\Delta_r v$ or the coarse-grained dissipation $\varepsilon_r$ as usually done \citep{Frisch.1995}. The motivation is two-fold. First, the observation of finite energy dissipation for infinite Reynolds numbers point to the existence of singularities in velocity gradients. Second, the existing method  to compute singularities is based on the gradients of the turbulent variable to overcome the influence of long-range correlations \citep{Turiel.2000}. Notice, however, that it can be build based on any variable whose singularity exponents are a linear combination of the singularity exponents of velocity.

Since the seminal work of \citet{Parisi.1985},  thirty-five years ago, a large theoretical body has been developed concerning the multifractal theory of turbulence (see the reviews by \citet{Boffetta.2008, Dubrulle.2019}
and class notes by \citet{Eyink.2019}). Nevertheless, the lack of an appropriate method to compute singularity exponents has constrained its application mainly to the analysis of statistical quantities such as the moments of given variables or structure functions \citep[e.g.][]{Verrier.2014, Bruyn-Kops.2015}. This has had two major consequences. First, some controversies, such as the validity of the Log-Normal or the Log-Poisson models, have remained open. Second, the geometrical properties of singularity exponents have not been analyzed with detail.  The publication of a fast and robust method to compute singularity exponents by \citet{Pont.2013} should contribute to change this situation. Indeed, this study is one of the firsts attempts to use the multifractal theory of turbulence based on the direct computation of singularity exponents to get insight about the energy cascade in the ocean.

In this study, we have defined simple shape parameters that allow to characterize and discriminate among cascade models. Of particular importance is the test on the symmetry of singularity spectra. Such a test is difficult to apply to the structure functions or the moments of a turbulent variable because it would require to compute the moments of negative order, that are usually ill-behaved. On the contrary, using an approach based on the statistics of singularity exponents, it is rather simple. In particular, the asymmetry of the observed singularity spectra (figure \ref{fig:singspec_properties}) implies that the log-Normal model should be discarded. Moreover, our results have shown that below the mixed layer the Log-Poisson is able to reproduce the observations, within the observational errors. Furthermore using the Log-Poisson model it has been possible to analytically derive the observed relation between the amplitude of the singularity spectrum and the most singular exponent. Although we cannot discard that other models may yet fit  the observations, particularly if better observations become available, the found evidence makes a strong case in favour of Log-Poisson model. 

A recent study based on a global ocean model proposes that the Log-Normal model is the most adequate based on the PDF of energy dissipation \citep{Pearson.2018}. This apparent discrepancy with the present work could be solved by the analysis of the singularity spectra of their numerical simulations. Indeed, making the approximation $\log\varepsilon_r\approx - Ah_\varepsilon + B$, where $A$ and $B$ are constants, it is possible to qualitatively reproduce the PDF of figure 1 in \citet{Pearson.2018} for the upper layers using equation \ref{eq:pdf-h}, the Log-Poisson model (equation \ref{eq:Log-Poisson-Dh}) and the values for $h_\infty$ and $\beta$ of figures \ref{fig:singspec_properties} and \ref{fig:logpois_assesment}. Moreover, in their study they observe non-zero skewness which is better in line with the Log-Poisson model instead of the Log-Normal. In addition, the analysis of the global model used by \citet{Pearson.2018}, or the model used by \citet{Su.2018}, should allow to investigate the geographical variability of our findings. Its worth mentioning that our multifractal approach is also very well suited to validate ocean models due to its scale invariance which allows to compare data and models of different spatial resolutions  \citep[e.g.][]{Skakala.2016}. 

As a final remark, the capability to compute the singularity exponents unveils the spatial structure of the energy cascade. As discussed in previous sections, singularity exponents give information about the local energy flux without the need to measure the velocity field below the resolution of observations. In particular, the spatial distribution of singularities shows that the smallest singularity exponents are located in the Gulf Stream and mesoscale vortices (e.g. figure \ref{fig:example-exps}B) thus linking large and small scale intermittency \citep{Feraco.2018, Pouquet.2019}.  Moreover, the analysis based on singularity exponents allows to easily segmentate the domain into sub-areas and explore the geographical dependence of energy dissipation pathways.  Interestingly, in situ observations have shown that three-dimensional turbulence is enhanced in ocean fronts \citep{dasaro.2011}. This pictures is apparently coherent with  the interpretation of singularity exponents as a measure of the intensity of the flux of energy towards small scales suggested by equation \ref{eq:scaling-Pi}.

The introduction of the scalings of velocity differences and coarse-grained dissipation (equations \ref{eq:h-definition-Du} and \ref{eq:h-definition-diss}) into Kolmogorov's Refined Similarity Hypothesis (equation \ref{eq:refined-similarity-hypothesis}) implies that $h$ has to be a constant, contradicting the observations here presented ($\Delta h^->0$), or $\nu$ should be a multifractal variable instead of a physical constant (at constant temperature).  On the contrary, \citet{Kraichnan.1974} argued that, in the inertial range, the local energy flux $\Pi_r(\vec{x})$ should be used instead, i.e. 
\begin{equation}
\Delta_ru \sim (r\Pi_r)^\frac{1}{3},
\label{eq:refined-similarity-hypothesis-PI}
\end{equation}
what is known as the Kraichnan's Refined Similarity Hypothesis. This  points out to the use of Kraichnan's Refined Similarity Hypothesis  (equation \ref{eq:refined-similarity-hypothesis-PI}) instead of Kolmogorov's (equation \ref{eq:refined-similarity-hypothesis}), so the central variable should be the local energy flux instead of energy dissipation.

\section{Conclusions}
\label{sec:conclusion}

The multifractal theory of turbulence has been used to investigate the vertical variation of the energy cascade in the North-Western Atlantic ocean. Contrary to the standard approach,  the focus of this study has been put on singularity exponents rather than on structure functions to analyze observations of turbulent flows. This approach has allowed to answer  to the symmetry conjecture in the ocean and to identify the Log-Poisson model as a better candidate to explain observations than the widely used Log-Normal model. Moreover, ocean observations have unveiled a  link between the anomalous scaling of the structure functions and dissipation, which could be also be derived analytically using the Log-Poisson model. It has been also shown that velocity statistics (anomalous scaling) depends on the strong velocity gradients associated to mesoscale vortices and the Gulf Stream.

\acknowledgments
We would like to thanks the insightful comments done by Emilio Garc\'{\i}a-Ladona and Joaquim Ballabrera-Poy. This work was supported by the Ministry of Economy and Competitiveness, Spain, and FEDER EU through the EXPLORA Ciencia  National R+D Plan under TURBOMIX (CGL2015-73100-EXP) project. We also acknowledge support from Fundaci\'on General CSIC (Programa ComFuturo).

%
%
\datastatement
Singularity exponents derived from velocity measurements can be downloaded from the PANGAEA repository
https://www.pangaea.de/tok/e30a52dc74ee28e36c19f15fefedd0b1c7372a1b (provisional link valid until November 10th 2020)

\appendix

Given the singularity spectrum $D(h)$ and the scaling of the moments $\tau(p)$ of $\overline{|\nabla u|}_r (\vec{x})$, it is possible to derive the corresponding functions for any other turbulent variable $\bar{\gamma}_r(\vec{x})$ related to $\overline{|\nabla u|}_r (\vec{x})$ by scaling relations of the form
\begin{equation}
\bar{\gamma}_r\sim r^b\overline{|\nabla u|}_r (\vec{x}) ^a
\label{sup-eq:scaling-epsilon-gamma}
\end{equation}
with
\begin{equation}
\langle \bar{\gamma}_r^p\rangle\sim r^{\tilde{\zeta}(p)}.
\label{sup-eq:gamma-mom}
\end{equation}
Indeed, from equations \ref{sup-eq:epsilon-mom}, \ref{sup-eq:scaling-epsilon-gamma} and \ref{sup-eq:gamma-mom} it can be shown that 
\begin{equation}
\tilde{\zeta}(p) = pb + \tau(ap),
\label{sup-eq:gamma-sing-zeta}
\end{equation}
which is also related to the singularity spectrum of this new variable $\tilde{D} (h)$ by a Legendre transform pair: 
\begin{equation}
\tilde{\zeta}(p)=\min_{h}[ph+d-\tilde{D}(h)],
\label{sup-eq:Zp-Legendre}
\end{equation}
\begin{equation}
\tilde{D}(h)=\min_{p}[ph+d-\tilde{\zeta}(p)].
\label{sup-eq:Dh-Legendre}
\end{equation}
Furthermore, the singularity exponents of the new variable $\tilde{h}(\vec{x})$ are related to those of $\overline{|\nabla u|}_r (\vec{x})$ by
\begin{equation}
\tilde{h}(\vec{x}) = b + a h(\vec{x}).
\label{sup-eq:gamma-sing}
\end{equation}

The scaling according to the Log-Poisson model for the variable $\bar{\gamma}_r(\vec{x})$ related to $\overline{|\nabla u|}_r (\vec{x})$ by a relation of the type given by equation \ref{sup-eq:scaling-epsilon-gamma} will be given by
\begin{equation}
\tilde{\zeta}(p) = pb + ah_\infty p + (d-D_\infty)(1 - \beta^{ap})
\end{equation}
and, using the Legendre transform (equation \ref{sup-eq:Dh-Legendre}), the corresponding singularity spectrum is
\begin{equation} 
\tilde{D}(h) =   D_\infty - \frac{h-ah_\infty-b}{\ln\beta^{a}}\left[1 - \ln\left(-\frac{h - ah_\infty-b}{(d - D_\infty)\ln\beta^{a}}\right)\right] .
\end{equation}
From the above definition it is possible the define two new parameters $\tilde{h}_{\infty}\equiv ah_\infty+b$ and $\tilde{\beta}\equiv \beta^a$ from which the original form of the Log-Poisson model (equation \ref{eq:Log-Poisson-Dh}) is recovered. Moreover, the relation between the anomalous scaling of the structure functions and the singularity exponents of $\bar{\gamma}_r(\vec{x})$ 
\begin{equation}
\left. \frac{d\zeta}{dp}\right|_{p=0} - \left. \frac{d\zeta}{dp}\right|_{p\rightarrow\infty} =  \frac{1}{a}\frac{\ln\beta}{1-\beta}(\min(\tilde{h})-b),
\label{eq:linear-model-Struct-general}
\end{equation}
notice that $\tilde{h}_{\infty}\equiv\min(\tilde{h})$.

 \bibliographystyle{ametsoc2014}
\bibliography{/Users/jisern/Recerca/Bibliografia/odc.bib}

%

%


\end{document}